# Wealth Share Analysis with "Fundamentalist/Chartist" Heterogeneous Agents

**Hai-Chuan Xu[1,2], Wei Zhang[1,2], Xiong Xiong[1,2*], Wei-Xing Zhou[3,4,5]**

[1]Collage of Management and Economics, Tianjin University, Tianjin, 300072, China;

[2]China Center for Social Computing and Analytics, Tianjin University, Tianjin, 300072, China

[3]School of Business, East China University of Science and Technology, Shanghai, 200237, China

[4]Department of Mathematics, East China University of Science and Technology, Shanghai, 200237, China

[5]Research Center for Econophysics, East China University of Science and Technology, Shanghai, 200237, China

**Abstract**：We build a multi-assets heterogeneous agents model with fundamentalists and chartists, who make investment decisions by maximizing the constant relative risk aversion utility function. We verify that the model can reproduce the main stylized facts in real markets, such as fat-tailed return distribution, long-term memory in volatility, and so on. Based on the calibrated model, we study the impacts of the key strategies' parameters on investors' wealth shares. We find that, as chartists' exponential moving average periods increase, their wealth shares also show an increasing trend. This means that higher memory length can help to improve their wealth shares. This effect saturates when the exponential moving average periods are sufficiently long. On the other hand, the mean reversion parameter has no obvious impacts on wealth shares of either type of traders. It suggests that, no matter fundamentalists take moderate strategy or aggressive strategy on the mistake of stock prices, it will has no different impact on their wealth shares in the long run.

## 1. Introduction

Compared with traditional economic modeling, agent-based modeling is more flexible in terms of characterizing the individual heterogeneity and population dynamics. This advantage is beneficial in researching on the survivability of different types of investors, namely market selection.

Previously, the related studies on market selection and wealth share distribution concentrate mainly on the impact of prediction accuracy, risk-aversion level, learning process, and noise trading. Blume and Easley [1] associated market selection with the first-principle of welfare economics, and discovered that in complete market under Pareto optimum allocation, the survival and disappearance of investors depend on the accuracy of their forecasts. Similarly, Fedyk, Heyerdahl and Walden [2] studied the multi-asset economy situation and found that, compared with rational investors, unsophisticated investors could suffer severe loss in the long run even their predication deviations seem priori small. Barucci and Casna [3] also found that under the mean reverting environment, investors who have inaccuracy predictions cannot survive.

Conversely, Chen and Huang [4, 5] compared the influence of forecasting accuracy and risk preference for investors' long-term survival, based on a multi-assets agent-based artificial stock market. They put forward that the risk preference was the determinant, and showed the wealth of the investors who adopted logarithmic utility function could be dominated in the long run. In the respect of learning evolution, LeBaron [6] focused on investors' learning on the gain level, which is the weight level for the last step's forecast error in their price forecast rules. This study showed the stylized facts of the market, analyzed the wealth evolution between different strategic investors and demonstrated their influence on the market instability. Amir, Evstigneev and Schenk [7] identified the adaptive portfolio strategies which could allow investors to survive under the frame of game theory. From the respective of noise trading, several researchers used the agent-based modeling method to analyze the survival problem in the long run by comparing the specialists and the noise traders, the noise traders & BSV investors, respectively [8, 9, 10]. Zhao [11] studied the survival boundary conditions of different irrational investors by utilizing the market utility maximization rather than the individual utility maximization.

---

[*] Corresponding author: xxpeter@tju.edu.cn (X. Xiong).

In the field of heterogeneous agents, fundamentalist/chartist modeling is a very important frame. For instance, Chiarella and He [12], Chiarella, Dieci and Gardini[13], Anufriev and Dindo[14], Yuan and Fu [15] and Zou and Ma [16] have mainly focused on the price equilibrium and system stability. This paper, however, focuses on the strategies parameters' impacts on investors' long-term wealth share, including fundamentalists' mean reversion parameters and chartists' exponential moving average periods under the fundamentalist/chartist modeling frame.

## 2 Heterogeneous agent model

For generality, this paper extends to multi-assets case based on the Constant Relative Risk Aversion (CRRA) heterogeneous agent model which was proposed by Chiarella, Dieci and Gardini [13]. The setting of the model is as follows.

### 2.1 The market

This paper proposes a discrete-time model with $n$ risky assets and one risk-free asset in the financial market. The risk-free interest rate $r$ is constant. There are two types of strategic agents, fundamentalists and chartists, and a Walrasian auctioneer.

Consider risky asset $i$. Under the assumption of traditional financial economics, investors have homogeneous rational expectations to the return of asset $i$, and the fundamental price of asset $i$ can be derived from the "no-arbitrage" equation

$$E_t[P_{i,t+1} + D_{i,t+1}] = (1+r)P_{i,t}. \tag{1}$$

The fundamental long-run solution is given by

$$P_{i,t} = P^*_{i,t} \equiv \sum_{k=1}^{\infty} \frac{E_t[D_{i,t+k}]}{(1+r)^k}, \tag{2}$$

where $P$, $P^*$, and $D$ denote the price, the fundamental value, and the dividend yield, respectively. In particular, if the dividend process is described by

$$E_t[D_{i,t+k}] = (1+\phi_i)^k D_{i,t}, k = 1, 2, ..., 0 \leq \phi_i \leq r, \tag{3}$$

one can obtain

$$P^*_{i,t} = (1+\phi_i)D_{i,t}/(r-\phi_i)$$

where $\phi_i$ denotes the dividend growth rate of asset $i$. Then one can easily obtain that the fundamental values evolve over time according to

$$E_t[P^*_{i,t+1}] = (1+\phi_i)P^*_{i,t} \tag{4}$$

and that the capital return is given by

$$E_t[\eta_{i,t+1}] \equiv E_t[\frac{P^*_{i,t+1}-P^*_{i,t}}{P^*_{i,t}}] = \phi_i \tag{5}$$

where $\eta_i$ denotes the dividend growth rate of asset $i$. In the following section we will introduce heterogeneity into the model. We assume agents have heterogeneous, time-varying beliefs about the first and second moment of capital returns, but for simplicity, they are assumed to share the same beliefs about dividend returns.

**2.2 The demand function**

Each agent is assumed to maximize the CRRA (power) utility function to allocate their wealth,

$$U^j(W) = \begin{cases} \frac{1}{1-\lambda^j}(W^{1-\lambda^j}-1) & (\lambda^j \neq 1), \\ \ln(W) & (\lambda^j = 1), \end{cases} \tag{6}$$

where $W>0$ represents the wealth and the parameter $\lambda^j>0$ represents the relative risk aversion coefficient. We choose the CRRA utility function because this assumption is quite realistic. The experiment results of Levy, Levy and Solomon [17] support the decreasing absolute risk aversion (DARA). In other words, , investor's risk aversion declines with the increase of wealth, which is consistent with the CRRA (power) utility function. In addition, Campbell and Viceira [18] pointed out that relative risk aversion cannot depend strongly on wealth in the long-run behavior of the economy.

This paper extends the solution proposed by Chiarella & He [12] to the multi-assets case and derives investors' demand function. At time $t$, the optimal wealth proportion $\pi_t$ to be invested in the risky asset is determined by maximizing the expected utility of wealth at $t+1$, as given by

$$\max_{\pi_t} E_t[U(W_{t+1})] \tag{7}$$

To solve this, one needs to work out the evolution of $U(W(t))$.

Assume that the wealth $W(t)$ follows a continuous time stochastic differential equation

$$dW = \mu(W)dt + \sigma(W)dz(t) \tag{8}$$

where $z(t)$ is a Wiener process. Let $X=U(W)$ be an invertible differentiable function with the inverse function $W=G(X)$. Following Ito's Lemma,

$$dX = [U'(W)\mu(W) + \frac{1}{2}\sigma^2(W)U''(W)]dt + \sigma(W)U'(W)dz \tag{9}$$

which can be written as

$$dX = \mu(X)dt + \sigma(X)dz \tag{10}$$

where

$$\mu(X) = U'(G(X))\mu(G(X)) + \frac{1}{2}\sigma^2(G(X))U''(G(X)) \tag{11}$$

and

$$\sigma(X) = \sigma(G(X))U'(G(X)) \tag{12}$$

Discretizing (10) using the Euler formula, one obtains the following approximation

$$X(t+\Delta t) = X(t) + \mu(X(t))\Delta t + \sigma(X(t))\Delta z(t) \tag{13}$$

It follows that

$$E_t[X(t+\Delta t)]=X(t)+\mu(X(t))\Delta t \tag{14}$$

$$V_t[X(t+\Delta t)]=\sigma^2(X(t)) \tag{15}$$

Unitizing the time $\Delta t$ in equation (14), we have

$$E_t[X_{t+1}]=X_t+\mu(X_t) \tag{16}$$

Substituting (11) into (16), one gets

$$E_t[U(W_{t+1})]\approx U(W_t)+\mu_t(W_t)U'(W_t)+\frac{1}{2}\sigma_t^2(W_t)U''(W_t) \tag{17}$$

This is the evolution of $U(W(t))$.

Let

$$\varphi_{i,t+1}=E_t(\varphi_{i,t+1})+\sigma_{i,t+1}\xi_{i,t} \tag{18}$$

where $\varphi_{i,t+1}$ is the return of asset $i$, $\sigma_{i,t+1}$ is the standard deviation of the return of asset $i$, and $\xi_{i,t}$ is an $N(0,1)$ process.

Meanwhile, we assume that a trader's wealth change equals to the sum of the return of the risk-free asset and the returns of risky assets, that is,

$$W_{t+1}-W_t=r(1-\sum_{i=1}^{n}\pi_{i,t})W_t+W_t\sum_{i=1}^{n}\pi_{i,t}\varphi_{i,t+1} \tag{19}$$

where $\pi_{i,t}$ is the wealth proportion invested in asset $i$ at time $t$.

Substituting (18) into (19), we have

$$\begin{aligned}W_{t+1}-W_t &= r(1-\sum_{i=1}^{n}\pi_{i,t})W_t+W_t(\sum_{i=1}^{n}\pi_{i,t}E_t(\varphi_{i,t+1})+\sum_{i=1}^{n}\pi_{i,t}\sigma_{i,t+1}\xi_{i,t})\\ &=[r(1-\sum_{i=1}^{n}\pi_{i,t})+\sum_{i=1}^{n}\pi_{i,t}E_t(\varphi_{i,t+1})]W_t+W_t\sum_{i=1}^{n}\pi_{i,t}\sigma_{i,t+1}\xi_{i,t}\end{aligned} \tag{20}$$

which can be written as

$$W_{t+1}-W_t=\mu_t(W)+\sigma_t(W)\xi_t \tag{21}$$

where

$$\mu_t(W)=[r(1-\sum_{i=1}^{n}\pi_{i,t})+\sum_{i=1}^{n}\pi_{i,t}E_t(\varphi_{i,t+1})]W_t=[r(1-\boldsymbol{\pi'_t}\mathbf{e})+\boldsymbol{\pi'_t}E_t(\boldsymbol{\varphi_{t+1}})]W_t \tag{22}$$

$$\begin{aligned}\sigma_t^2(W) &= D(W_t\sum_{i=1}^{n}\pi_{i,t}\sigma_{i,t+1}\xi_{i,t})\\ &=W_t^2[\sum_{i=1}^{n}\pi_{i,t}^2\sigma_{i,t+1}^2 D(\xi_{i,t})+\sum_{i=1}^{n}\sum_{j=1}^{n}\rho_{ij}\pi_{i,t}\pi_{j,t}\sigma_{i,t+1}\sigma_{j,t+1}\sqrt{D(\xi_{i,t})}\sqrt{D(\xi_{j,t})}]\\ &=W_t^2[\sum_{i=1}^{n}\pi_{i,t}^2\sigma_{i,t+1}^2+\sum_{i=1}^{n}\sum_{j=1}^{n}\rho_{ij}\pi_{i,t}\pi_{j,t}\sigma_{i,t+1}\sigma_{j,t+1}]\\ &=W_t^2[\sum_{i=1}^{n}\pi_{i,t}^2\sigma_{i,t+1}^2+\sum_{i=1}^{n}\sum_{j=1}^{n}\pi_{i,t}\pi_{j,t}\sigma_{ij,t+1}]\\ &=W_t^2\boldsymbol{\pi'_t}\Sigma_t\boldsymbol{\pi_t}\end{aligned} \tag{23}$$

with

$$\boldsymbol{\pi_t}=[\pi_{1,t},\pi_{2,t},\cdots\pi_{n,t}]'$$

and

$$\Sigma_t=\begin{pmatrix}\sigma_{1,t+1}^2 & \cdots & \sigma_{1n,t+1}\\ \vdots & \ddots & \vdots\\ \sigma_{n1,t+1} & \cdots & \sigma_{n,t+1}^2\end{pmatrix}$$

Substituting (22) and (23) into (17), we have

$$E_t[U(W_{t+1})] \approx U(W_t) + [r(1-\pi'_t \mathbf{e}) + \pi'_t E_t(\varphi_{t+1})] W_t U'(W_t) + \frac{1}{2} W_t^2 \pi'_t \Sigma_t \pi_t U''(W_t) \quad (24)$$

Thus the first order condition of the problem (7) leads to the following optimal solution

$$\pi_t = -\frac{U'(W_t)}{W_t U''(W_t)} \Sigma_t^{-1} [E_t(\varphi_{t+1}) - r\mathbf{e}] = \frac{1}{\lambda} \Sigma_t^{-1} [E_t(\varphi_{t+1}) - r\mathbf{e}] \quad (25)$$

where $E_t(\varphi_{t+1})-r\mathbf{e}$ represents the vector of the expected excess return on risky assets, $\Sigma_t$ represents the covariance matrix of the expected return on risky assets. Then we can get the investor's position demand for all assets:

$$\mathbf{z_t} = W_{t-1} \pi_t ./ \mathbf{P_t} \quad (26)$$

where $\mathbf{P_t}$, $\mathbf{z_t}$, and ./ denote the price vector, the vector of the demand for asset position, and the division of the element as opposed to the vector, respectively. It can be seen that, although the optimal investment proportion of an investor's wealth to be invested in the risky asset is independent of wealth, its optimal position demand is proportional to wealth.

**2.3 Heterogeneous expectations**

The heterogeneous beliefs of fundamentalists and chartists are reflected in the expected return as well as the expected variance. We assume that the same type of investors predict all risky assets in the same way

*2.3.1 Fundamentalists*

Assume that fundamentalists (denoted by $f$) know the fundamental value of assets. These investors believe that the price will move back to the fundamental value when the market price deviates from fundamental value. Therefore their expected price change is

$$\begin{aligned} E_t^{(f)}[P_{i,t+1} - P_{i,t}] &= E_t^{(f)}[P_{i,t+1}^* - P_{i,t}^*] + d_f(P_{i,t}^* - P_{i,t}) \\ &= \phi P_{i,t}^* + d_f(P_{i,t}^* - P_{i,t}) \end{aligned} \quad (27)$$

Fundamentalists' expected price is the sum of the change of the fundamental value and an adjustment item. The adjustment item is proportional to the deviation between current asset price and the fundamental value. The coefficient $d_f$ ($d_f > 0$) indicates the speed returning to the fundamental value, and we call it mean reversion coefficient. In addition, to simplify, we assume that the two types of investors have the same expectation to dividend return, that is

$$E_t(D_{i,t+1}) = E_t^{(f)}(D_{i,t+1}) = E_t^{(c)}(D_{i,t+1}) = (1+\phi) D_{i,t} \quad (28)$$

Thus, fundamentalist's expected return for asset $i$ is

$$\begin{aligned} E_t^{(f)}(\varphi_{i,t+1}) &= \frac{E_t^{(f)}[P_{i,t+1}] + E_t^{(f)}(D_{i,t+1}) - P_{i,t}}{P_{i,t}} \\ &= \frac{\phi P_{i,t}^* + d_f(P_{i,t}^* - P_{i,t}) + (1+\phi) D_{i,t} - P_{i,t}}{P_{i,t}} \end{aligned} \quad (29)$$

We further assume that fundamentalists use the exponential moving average of expected return deviation to determine the expected return variance. To certain extent, it reflects the adaptability, i.e.

$$_t\sigma^2_{i,t+1} = e^{-1/\tau^{(f)}}\,_{t-1}\sigma^2_{i,t} + (1-e^{-1/\tau^{(f)}})(E^{(f)}_{t-1}[\varphi_{i,t}]-\varphi_{i,t})^2 \tag{30}$$

where $\tau^{(f)}$ stands for the period length of the exponential moving average. The larger is $\tau^{(f)}$, the smaller is $(1-e^{-1/\tau^{(f)}})$. It shows that the longer is the moving average period, the smaller is the weight of the latest deviation. Thus the latest expected deviation has the smaller influence on the expected variance. Furthermore, this paper assumes that investors' expected correlations between assets is $\rho^{(f)}_{ij}$ ($i,j$ denotes the assets), which varies with different investors, but does not change with time. Hence, the covariance matrix of the expected return is

$$\Sigma^{(f)}_t = \begin{pmatrix} _t\sigma^2_{1,t+1} & \cdots & \rho^{(f)}_{1n}\,_t\sigma_{1,t+1}\,_t\sigma_{n,t+1} \\ \vdots & \ddots & \vdots \\ \rho^{(f)}_{n1}\,_t\sigma_{n,t+1}\,_t\sigma_{1,t+1} & \cdots & _t\sigma^2_{n,t+1} \end{pmatrix} \tag{31}$$

*2.3.2 Chartists*

Chartists do not know the fundamental value of the assets. They use the past price series to infer the movement of future prices. Therefore, chartists can be regarded as a kind of adaptive investors. This paper assumes that the first moment and the second moment of the expected return are adaptable for chartists, i.e. both the expected return and its variance are obtained by using the exponential moving average. The expected price change of chartists is

$$\begin{aligned} m^{(c)}_t &\equiv E^{(c)}_t[\eta_{t+1}] = E^{(c)}_t[\frac{P_{t+1}-P_t}{P_t}] \\ &= e^{-1/\tau^{(c)}} m^{(c)}_{t-1} + (1-e^{-1/\tau^{(c)}})(\frac{P_t-P_{t-1}}{P_{t-1}}) \end{aligned} \tag{32}$$

where $\tau^{(c)}$ presents the period length of the exponential moving average. Chartists have the same expected dividend yield as fundamentalists, whose expected return for asset $i$ is

$$\begin{aligned} E^{(c)}_t(\varphi_{i,t+1}) &= \frac{E^{(c)}_t[P_{i,t+1}]+E^{(c)}_t(D_{i,t+1})-P_{i,t}}{P_{i,t}} \\ &= e^{-1/\tau^{(c)}} m^{(c)}_{t-1} + (1-e^{-1/\tau^{(c)}})(\frac{P_t-P_{t-1}}{P_{t-1}}) + \frac{(1+\phi)D_{i,t}}{P_{i,t}} \end{aligned} \tag{33}$$

For chartists, the covariance matrix of the expected return is similar to that of fundamentalists.

**2.4 Market clearing**

This model achieves market clearing through the equilibrium between supply and demand as

$$\sum_{j=1}^{N^{(f)}} \mathbf{Z}^j + \sum_{j=N^{(f)}+1}^{N} \mathbf{Z}^j = \mathbf{M} \tag{34}$$

where $j$ represents the investors, $N^{(f)}$ represents the number of fundamentalists, $N$ is the total number of investors, $N-N^{(f)}=N^{(c)}$ is the number of chartists, and $\mathbf{M}$ is the number of outstanding stocks in the market. The equation indicates that the sum of risky asset holdings by all traders are equal to $\mathbf{M}$ at any time $t$, which is achieved by adjusting the equilibrium price repeatedly.

**2.5 Wealth shares**

After the market price is determined through the clearing mechanism, the wealth of investor $j$ is also determined:

$$W_t^j = (1 - \sum_{i=1}^{n} \pi_{i,t})W_{t-1}(1+r) + W_{t-1}\sum_{i=1}^{n} \pi_{i,t}\frac{P_{i,t}+D_{i,t}}{P_{i,t-1}} \tag{35}$$

At this moment, the total wealth of all investors, the fundamentalists and the chartists in the market are

$$W_t = \sum_{j=1}^{N} W_t^j$$

$$W_t^{(f)} = \sum_{j=1}^{N^{(f)}} W_t^j \tag{36}$$

$$W_t^{(c)} = \sum_{j=1}^{N^{(c)}} W_t^j$$

where $N$ is the number of investors, $N^{(f)}$ is the number of fundamentalists, $N^{(c)}$ is the number of chartists, $W_t^{(f)}$ is the total wealth of fundamentalists, and $W_t^{(c)}$ is the total wealth of chartists. Accordingly, investor $j$'s wealth share, all fundamentalists' wealth share and chartists' wealth share in the market are defined as follows

$$w_t^j = \frac{W_t^j}{W_t}$$

$$w_t^{(f)} = \frac{W_t^{(f)}}{W_t} \tag{37}$$

$$w_t^{(c)} = \frac{W_t^{(c)}}{W_t}$$

By studying the relative wealth share rather than the absolute wealth amount, we can get to the wealth evolution of different types of investors more intuitively.

## 3. Simulation results

To reflect the randomness of the dividend process, we revised the dividend process in the following agent-based experiments as

$$D_{i,t+1} = (1 + \phi_i + \sigma_{i,\varepsilon}\varepsilon_{t+1})D_{i,t} \tag{38}$$

Then we can obtain the fundamental value of the asset

$$P_{i,t+1}^* = (1 + \phi_i + \sigma_{i,\varepsilon}\varepsilon_{t+1})P_{i,t}^* \tag{39}$$

where $\varepsilon_t \sim N(0,1)$, $\sigma_{i,\varepsilon} > 0$ represents standard deviation dividend growth rates.

**3.1 Reproducing stylized facts**

Table 1 lists the parameters for the benchmark model. In this paper, we use three risky stocks as examples to calibrate this model and the parameters of all stocks are setting consistently. The fundamental value of each stock is setting to 10, each investor' initial wealth is 10, so that his total initial wealth is 40. In addition, short selling is permitted in the model. The total number of investors is 40, including 20 fundamentalists and 20 chartists. One step in this model can be seen as one week in reality. Every experiment runs 1000 periods, corresponding 20 years in reality.

The risk-free interest rate is 0.0012, corresponding to the annual interest rate which is about 6%. The dividend growth rate is 0.001, corresponding to the annual growth rate is about 5%The initial dividend is 0.002. Many studies suggest that the relative risk aversion is in the range from 2 to 4. In this paper, we set it to 3. To reflect the heterogeneity of the investors, the exponential moving average periods, fundamentalists' mean reversion parameter and the expected correlation coefficients between assets are randomly selected in a certain range by every trader at the beginning of the experiment, which are kept unchanged during the remaining experiment times. For example, the correlation coefficients between stocks are selected randomly in the range [-0.2, 0.8], which is consistent with real stock market.[1]

It is easily understand that, although stocks have the same parameters, due to the randomness of the dividend process, and the different imbalances of supply and demand, the price evolutions of different stocks are not the same, and they can even be opposite.

Table 1 The parameters of basic model

| Parameter | Value | Parameter | Value |
| --- | --- | --- | --- |
| No. of risky assets | 3 | Random seeds | 0 |
| No. of agents | 40 | The risk-free interest rate | 0.0012 |
| No. of Fundamentalist | 20 | The relative risk aversion | 3 |
| No. of Chartist | 20 | The max exponential moving average periods | 80 |
| Initial cash | 10 | The min exponential moving average periods | 20 |
| Initial stock positions | 1 | The min mean reversion parameter | 0.5 |
| The minimum stock positions | -5 | The max mean reversion parameter | 1 |
| The maximum stock positions | 10 | The min expected correlation coefficient | -0.2 |
| The initial dividend | 0.002 | The max expected correlation coefficient | 0.8 |
| Dividend growth rate | 0.001 | The max wealth investment proportion | 0.95 |
| The standard deviation of dividend growth rate | 0.01 | The min wealth investment proportion | -0.95 |

Considering that our model is a growth model, in which the dividend growth rate is positive, if we compare one day as one time step, then the dividend growth rate will be so small that the model accuracy could be lost. Thus, in this paper, we use weekly closing prices of the S&P 500 index from December 30, 1991 to March 7, 2011 as the calibration series, and compare it with the simulated price series in both the descriptive statistics and the stylized facts.

Figure 1 shows the evolution of stock prices. The bottom right plot is for the S&P 500 index. Table 2 shows the descriptive statistics of the three simulated stock prices and the S&P 500 prices. We can find that these properties of simulated prices are very similar with those of S&P 500 index.

---

[1] Ochiai and Nacher [19] show that the correlations between DJIA and Nikkei 225 roughly fluctuate within [-0.2, 0.8].

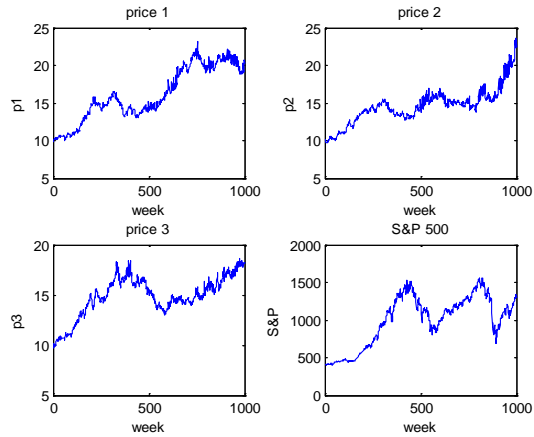

Figure 1 Evolution of prices of the three simulated stocks and the S&P 500 index.

Table 2 The descriptive statistics

| statistic | asset1 | asset2 | asset3 | S&P500 |
|---|---|---|---|---|
| mean | 0.00090 | 0.0011 | 0.00072 | 0.0014 |
| median | 0.0016 | 0.00048 | 0.0011 | 0.0024 |
| S. D. | 0.0196 | 0.0235 | 0.0157 | 0.0238 |
| kurtosis | 7.0397 | 7.9133 | 6.2793 | 8.9414 |
| skewness | 0.0496 | 0.1523 | 0.1301 | -0.5191 |

Here we test the stylized facts of our model. Figure 2-5 show the stylized facts of S&P 500, stock 1, stock 2, and stock 3. Each figure contains three plots. The top panel is the distribution of return rate, where the red line is the probability density curve of normal distribution. The middle panel is the autocorrelation function of the return rate. The bottom panel is the autocorrelation function of the absolute return rate. Obviously, the S&P500 returns exhibit fat tails. From the autocorrelation plots, we can find the original return has no autocorrelations, whereas the absolute return shows significant long-term autocorrelations. In addition, the Hurst indexes for the absolute returns of three simulated stocks are 0.6993, 0.7585 and 0.6817 respectively, which confirm the property of long-term autocorrelation as in real markets [20]. We find all three stocks show the similar characteristics as the S&P 500.

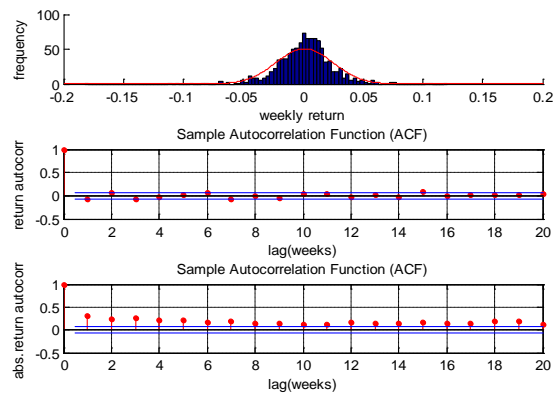

Figure 2 Main stylized fact of the S&P 500 index. The top panel shows the distribution of returns. The middle panel shows the autocorrelation function of the returns. The bottom panel is the autocorrelation of the absolute returns.

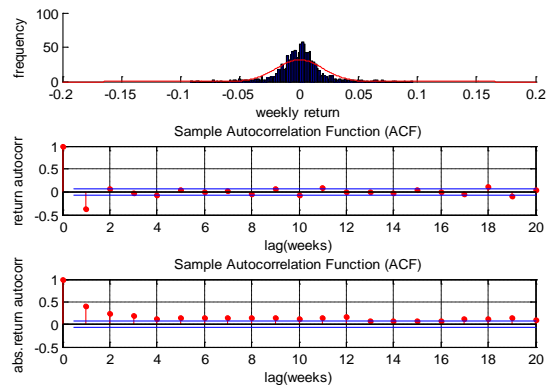

Figure 3 Main stylized fact of the simulated Stock 1. The top panel shows the distribution of returns. The middle panel shows the autocorrelation function of the returns. The bottom panel is the autocorrelation of the absolute returns.

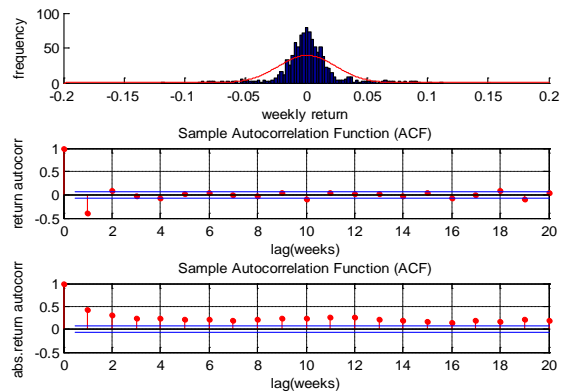

Figure4  Main stylized fact of the simulated Stock 2. The top panel shows the distribution of returns. The middle panel shows the autocorrelation function of the returns. The bottom panel is the autocorrelation of the absolute returns.

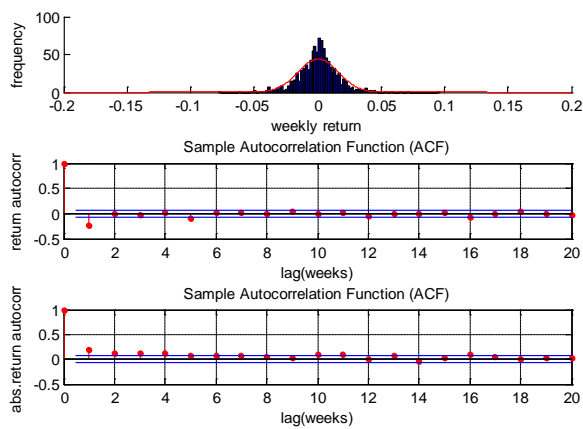

Figure 5 Main stylized fact of the simulated Stock 3. The top panel shows the distribution of returns. The middle

panel shows the autocorrelation function of the returns. The bottom panel is the autocorrelation of the absolute returns.

We conclude that our model is able to reproduce the main stylized facts of real stocks and stock indexes, including the fat-tailed distribution of returns, the absence of long-memory in the returns, and the strong long-term correlations in the absolute returns. It indicates that our model has captured some key ingredients of the microstructure of real financial markets.

### 3.2 Wealth share analysis

Investors' believes play an important role in making investment decisions. Therefore, it's essential to analyze the key parameters that determine investor's believes and focus on these parameters' impacts on investors' wealth accumulation. The model has two key parameters, including fundamentalists' mean reversion parameters $d_f$ and chartists' exponential moving average periods $\tau(c)$.

In order to analyze the two parameters' impacts on two types of investors' wealth shares, this paper divides the mean reversion parameter values into four intervals: (0.5,0.6), (0.6,0.7), (0.7,0.8) and (0.8,0.9). Fundamentalists' mean reversion parameters value is randomly selected in the specific interval. We also take 9 different values for the chartists, that is, 1,2,3,4,5,10,20,50, and 80, which correspond to 9 different weights of the latest price information, that is, 0.6321,0.3935,0.2835,0.2212,0.1813,0.0952,0.0488,0.0198, and 0.0124. This results in 36 different parameters combinations. In order to keep the conclusion robust, this paper selects five different random seeds for each combination to conduct five experiments, and then average investors' wealth shares of these five experiments. Thus, we run a total of 180 experiments. We focus on the aggregate wealth of the same type of traders. Figures 6 - 9 show investors' wealth share for 4 different intervals of $d_f$. In every figure, each bar corresponds to the average wealth share at the end of experiments when exponential moving average period $\tau(c)$ is given a specific value.

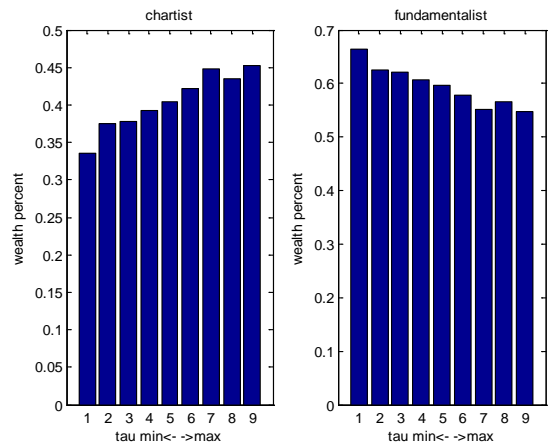

Figure 6 Wealth shares ($d_f \in (0.5,0.6)$)

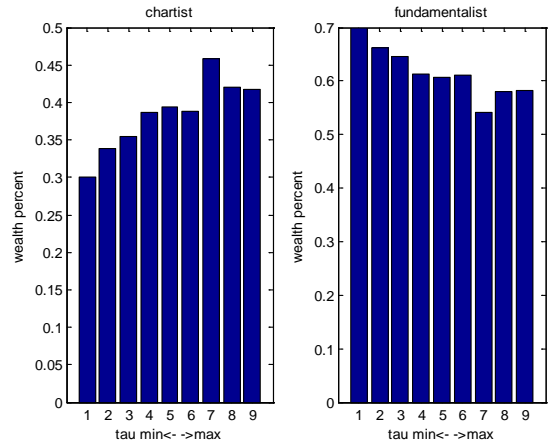

Figure 7. Wealth Shares $(d_f \in (0.6, 0.7))$

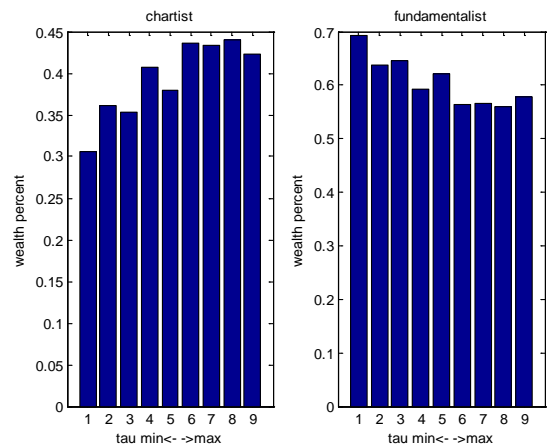

Figure 8. Wealth Shares $(d_f \in (0.7, 0.8))$

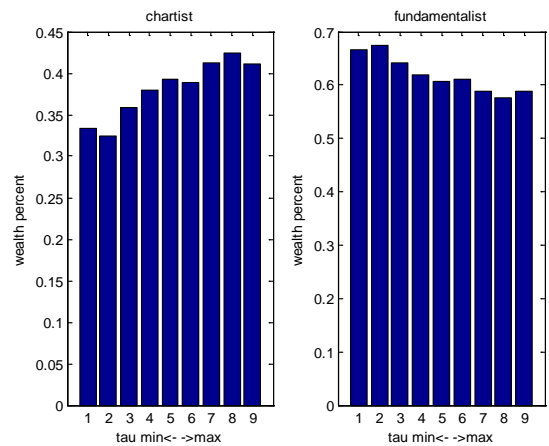

Figure 9. Wealth Shares $(d_f \in (0.8, 0.9))$

We can see that no matter what kinds of parameters' portfolios are chosen, the two types of investors co-exist in the long term. Chartists' wealth shares fall into the range of 0.3-0.5, and the corresponding fundamentalists' wealth shares locate between 0.7 and 0.5. Meanwhile, different portfolios $(d_f, \tau^{(c)})$ do have different distributions of wealth. Obviously, no matter $d_f$ is at which intervals, with the exponential moving average periods increasing, chartists' wealth shares also show an increasing trend, indicating that a higher memory length will help chartists form more accurate expectations, thus increasing their wealth shares. However, this trend becomes less obvious when exponential moving average periods are high enough, such as $\tau^{(c)}$ is 20、50、80. This is because chartists' return expectations only have very small weight on the latest price information (namely 0.0488, 0.0198 and 0.0124), and more than 95% of the weights are given to the past pricing information. Hence, the growth trend of wealth share is no longer obvious when the exponential moving average period is sufficiently long.

In addition, the value of mean reversion coefficient $d_f$ has no significant impacts on the wealth share of the two types of investors. It suggests that the aggressiveness of fundamentalists' strategies on the mistake of stock prices, be they moderate (when $d_f$ is small) or aggressive (when $d_f$ is large), has no impact on their wealth shares in the long run.

## 4. Conclusions

We have built a multi-assets heterogeneous-agents model with fundamentalists and chartists. We verified that the model can reproduce the main stylized facts in real markets such as fat tails in the return distribution, absence of long-memory in returns, long-term memory in the absolute returns, and so on. Based on the calibrated model, we studied the key strategies parameters' impacts on investors' wealth shares. We found that as chartists' exponential moving average periods increase, their wealth shares also show an increasing trend. This means that higher memory length can help to improve their wealth shares. However when the exponential moving average periods has been long enough, this trend is no longer obvious and no matter how long the memory is, wealth share of chartists will not be higher than fundamentalists'. That is, chartists' wealth share will not be more than 0.5. This reflects that chartists can co-exist with fundamentalists in stock markets, i.e. at least accounting for about 30% market wealth, although they cannot equally share the market wealth. On the other hand, the mean reversing parameter has no significant impacts on the wealth share of either type of traders. Therefore, no matter fundamentalists take moderate strategy or aggressive strategy on the mistake of stock prices, it has no different impacts on their wealth shares in the long run.

## Conflict of Interests

The authors declare that there is no conflict of interests regarding the publication of this paper.

## Acknowledgment

This research is partly supported by NSFC (No. 71131007, 71271145), Program for Changjiang Scholars and Innovative Research Team (No. IRT1028), the Ph.D. Programs Foundation of Ministry of Education of China (No. 20110032110031).